\documentclass[sigconf,nonacm]{acmart}

\DeclareGraphicsExtensions{.pdf,.png}

\usepackage{amsmath,amssymb,amsfonts}
\usepackage{graphicx}
\usepackage{subcaption}
\usepackage{textcomp}
\usepackage{xcolor}
\usepackage{booktabs}
\usepackage{listings}
\usepackage{placeins}
\usepackage{multicol} 
\lstdefinelanguage{proto}
  {
    morekeywords={int32,uint32,int64,message,bool},
    sensitive=false,
    morecomment=[l]{//},
    morecomment=[s]{/*}{*/},
    morestring=[b]",
  }
\lstdefinelanguage{proto-instance}
  {
    morekeywords={toy_id,in_inventory,price},
    sensitive=false,
    morecomment=[l]{//},
    morecomment=[s]{/*}{*/},
    deletestring=[b]",
  }
\lstdefinelanguage{text}
  {
    sensitive=true,
    morestring=[b]",
  }
\usepackage{xcolor}
\usepackage{color}
\usepackage{tabulary}
\usepackage{quoting}
\quotingsetup{leftmargin=9pt,vskip=7pt}
\usepackage{algorithm}
\usepackage{algpseudocode}
\definecolor{codegreen}{rgb}{0.06,0.57,0.11}
\definecolor{codegray}{rgb}{0.4,0.4,0.4}
\definecolor{codepurple}{rgb}{0.72,0,0.72}
\lstdefinestyle{mystyle}{
    commentstyle=\color{codegreen},
    numberstyle=\color{codegray},
    keywordstyle=\color{codepurple},
    basicstyle=\ttfamily\footnotesize,
    breakatwhitespace=false,
    breaklines=true,
    captionpos=b,
    keepspaces=true,
    numbers=left,
    numbersep=5pt,
    showspaces=false,
    showstringspaces=false,
    showtabs=false,
    tabsize=2
}
\usepackage{graphicx}
\graphicspath{ {./img/} }
\usepackage{siunitx}
\usepackage{balance}

\newcommand{\GoogleUSA}{Google}
\newcommand{\GoogleMunich}{Google}
\newcommand{\TotalIds}{39}
\newcommand{\TotalMonths}{twelve}
\newcommand{\TotalCls}{595}
\newcommand{\TotalClsDevs}{3}
\newcommand{\TotalClsDevsText}{three}
\newcommand{\TotalClsDevOne}{290}
\newcommand{\TotalClsDevTwo}{251}
\newcommand{\TotalClsDevThree}{54}
\newcommand{\TotalClsLlmOnly}{214}
\newcommand{\TotalClsLlmOnlyPct}{$35.97\%$}
\newcommand{\TotalClsLlmThenHuman}{229}
\newcommand{\TotalClsLlmThenHumanPct}{$38.48\%$}
\newcommand{\TotalClsHumanOnly}{152}
\newcommand{\TotalClsHumanOnlyPct}{$25.55\%$}

\newcommand{\TotalClsLlmPct}{$74.45\%$}
\newcommand{\TotalClReviewers}{$306$}
\newcommand{\TotalClTeams}{$149$}
\newcommand{\TotalClOffices}{$37$}
\newcommand{\TotalClTimezones}{$12$}
\newcommand{\TotalEdits}{$93,574$}
\newcommand{\TotalEditsLlm}{$64,996$}
\newcommand{\TotalEditsLlmPct}{$69.46\%$}
\newcommand{\TotalEditsHuman}{$28,578$}
\newcommand{\TotalEditsHumanPct}{$30.54\%$}
\newcommand{\TotalTimeReduction}{$50\%$}
\newcommand{\ToyId}{\texttt{toy\_id}}
\newcommand{\DELTA}{$\Delta$}
\newcommand{\TotalDelta}{Total\DELTA{}}
\newcommand{\LlmDelta}{\texttt{LLM\DELTA{}}}
\newcommand{\HumanDelta}{\texttt{Human\DELTA{}}}
\newcommand{\Leven}{Levenshtein}

\begin{document}

\title{Migrating Code At Scale With LLMs At Google}

\author{Celal Ziftci}
\authornote{All authors contributed equally to this research.}
\email{celal@google.com}
\affiliation{
  \institution{\GoogleUSA{}}
  \city{New York}
  \state{NY}
  \country{USA}
}

\author{Stoyan Nikolov}
\authornotemark[1]
\email{stoyannk@google.com}
\affiliation{
  \institution{\GoogleMunich{}}
  \city{Munich}
  \country{Germany}
}

\author{Anna Sjövall}
\authornotemark[1]
\email{annaps@google.com}
\affiliation{
  \institution{\GoogleMunich{}}
  \city{Munich}
  \country{Germany}
}

\author{Bo Kim}
\authornotemark[1]
\email{bohyungk@google.com}
\affiliation{%
  \institution{\GoogleUSA{}}
  \city{New York}
  \state{NY}
  \country{USA}
}

\author{Daniele Codecasa}
\authornotemark[1]
\email{cdcs@google.com}
\affiliation{
  \institution{\GoogleMunich{}}
  \city{Munich}
  \country{Germany}
}

\author{Max Kim}
\authornotemark[1]
\email{maxkim@google.com}
\affiliation{
  \institution{\GoogleUSA{}}
  \city{Mountain View}
  \state{CA}
  \country{USA}
}

\begin{abstract}

Developers often evolve an existing software system by making internal changes, called \textit{migration}. Moving to a new framework, changing implementation to improve efficiency, and upgrading a dependency to its latest version are examples of migrations.

Migration is a common and typically continuous maintenance task undertaken either manually or through tooling. Certain migrations are labor intensive and costly, developers do not find the required work rewarding, and they may take years to complete. Hence, automation is preferred for such migrations.

In this paper, we discuss a large-scale, costly and traditionally manual migration project at Google, propose a novel automated algorithm that uses change location discovery and a Large Language Model (LLM) to aid developers conduct the migration, report the results of a large case study, and discuss lessons learned.

Our case study on \TotalIds{} distinct migrations undertaken by \TotalClsDevsText{} developers over \TotalMonths{} months shows that a total of \TotalCls{} code changes with \TotalEdits{} edits have been submitted, where \TotalClsLlmPct{} of the code changes and \TotalEditsLlmPct{} of the edits were generated by the LLM. The developers reported high satisfaction with the automated tooling, and estimated a \TotalTimeReduction{} reduction on the total time spent on the migration compared to earlier manual migrations.

Our results suggest that our automated, LLM-assisted workflow can serve as a model for similar initiatives.

\end{abstract}

%


\maketitle

\section*{Preprint Notice}
This is a preprint of a paper accepted at the ACM International Conference on the Foundations of Software Engineering (FSE) 2025.

\newpage
\section{Introduction}

Migration is a crucial process in software development that involves modernizing and adapting systems to ensure ongoing functionality and compatibility. This may include, but is not limited to, updating or restructuring existing code, improving its organization, making it easier to understand and maintain, transitioning to newer versions of programming languages to leverage new features and address potential security vulnerabilities, updating dependencies such as APIs and libraries to modernize the codebase, fix bugs and improve performance, and adapting to changes in underlying frameworks and platforms that applications depend on.

Performing such changes manually, particularly in large codebases, is often tedious, time-consuming, and error-prone, especially when dealing with complex code structures and dependencies, resulting in barriers for developers \cite{hora2015developers, kula2018developers, wang2020empirical}. As a result, automating code migrations is critical in maintaining systems to avoid recurring manual work, since systems typically undergo several migrations over their lifetime.

Migrations are typically automated using different techniques, ranging from regular expression search and replace, to using code transformation tools, to specialized domain specific tooling. Existing techniques have shortcomings in their accuracy on finding relevant locations to change and the need for maintenance of the rules or the domain specific tools.

Recently, the introduction of Large Language Models (LLMs) created opportunities to support several software engineering tasks, including generating tests \cite{alshahwan2024automated, schafer2023empirical, siddiq2024using}, code refactoring \cite{shirafuji2023refactoring}, documentation generation \cite{dvivedi2024comparative}, bug detection \cite{alrashedy2023language}, fault localization \cite{kang2024quantitative}, fixing bugs \cite{sobania2023analysis}, code reviews \cite{tufano2021towards, tufano2022using}, pair programming \cite{imai2022github}, and recently migrations \cite{almeida2024automatic, omidvar2024evaluating}.

In this paper, we discuss an automated solution for a large, costly migration at Google on modifying 32-bit integers to 64-bits by automatically identifying areas of code that require modification, using LLMs to obtain the necessary code changes, running several forms of validation to confirm that the code changes are valid, and sending the changes to developers for verification and acceptance to enable more efficient, reliable and verifiable code migrations that use natural language to describe the migration.

To the best of our knowledge, our work is the largest, most comprehensive code migration study using LLMs to date. We assess the success of our solution on \TotalIds{} instances of this migration undertaken over \TotalMonths{} months, and report our learnings on the experiences of the \TotalClsDevsText{} developers that conducted the migrations.

\section{Migrating Identifiers From 32-bits To 64-bits}

In this section, we discuss the specific type of migration we conducted at Google, its various characteristics, and show-case such a migration on an example.

\subsection{Motivating Example}
At Google, a typical way to describe the inputs and outputs of a system is to use a protocol buffer \cite{protobuffers}, \textit{proto} in short, a structured data format that supports a wide range of scalar value types including enums, strings, integers and floating point numbers, as well as custom user-defined structured message formats built upon those scalar value types. Based on the proto definition, the compiler generates standardized library code in the user's target programming language of choice to manipulate instances of the proto.

\begin{lstlisting}[label={lst:proto-schema}, float, language=proto, numbers=none,
caption=Example proto schema to describe toys in a toy store's inventory.]
// A toy type known to our systems
message Toy {
  int32 toy_id;
  uint32 price; // In US dollars
  bool in_inventory;
}
\end{lstlisting}

Consider the proto schema in Listing \ref{lst:proto-schema} that describes different types of toys in an online toy store's inventory. \texttt{Toy} is a user-defined message type that contains three scalar fields: \ToyId{} is a 32 bit integer to uniquely identify a toy, \texttt{price} is a number that represents the price of the toy in US dollars, and \texttt{in\_inventory} is a boolean that represents whether a toy is currently in the store's inventory.

The system's inventory stores such proto instances in a database to show toy listings on a website. The system codebase in Java has references to the fields defined in the proto schema, with sample snippets shown in Listing \ref{lst:working-code} where there is a view that shows the toy with a different badge depending on when it was produced, and a test class with two test methods testing the view for toys that are produced before and after 2024.

Note that there are direct and indirect references to \ToyId{} in Listing \ref{lst:working-code}. Lines 5, 13, 27, 39 are direct references, since they directly refer to the access methods for \ToyId{} in the proto, while lines 36 and 37 are indirect references, since they do not directly refer to \ToyId{} but contribute to the value used in the direct reference on line 39 through data flow.

\begin{lstlisting}[label={lst:working-code}, float, language=Java, caption=Example code and tests in Java that references various fields of the proto schema.]
final class ToyView {    /* ToyView.java */
    public static HtmlView showToy(Toy toy) {
        if (toyIsProducedAfter2024(toy)) {
            // Create panel with "new" badge
            final int toyId = toy.getToyId();
            ...
        } else {
            // Return the default panel
        }
    }
    
    static boolean toyIsProducedAfter2024(Toy toy) {
        return firstBitIsOne(toy.getToyId(), 31));
    }
    
    // First ID bit is 1 if produced after 2024
    static boolean firstBitIsOne(long id, int shift) {
        return (id >> shift) == 1;
    }
}

final class ToyViewTest {    /* ToyViewTest.java */
  private static final Random R = new Random();
  
  public static int testShowToy_producedBefore2024() {
    final Toy testToy = Toy.Builder
                            .setToyId(-5)
                            .setPrice(12)
                            .IsInInventory(true).build();
    final HtmlView view = ToyView.showToy(testToy);
    // Assert that view has toy panel with "new" badge
    ...
  }
  
  public static int testShowToy_producedAfter2024() {
    final int randomId = R.nextInt(0, 100);
    final int toyId = randomId * 10;
    final Toy testToy = Toy.Builder
                            .setToyId(toyId)
                            .setPrice(23)
                            .IsInInventory(false).build();
    final HtmlView view = ToyView.showToy(testToy);
    // Assert that view has default panel
    ...
  }
}
\end{lstlisting}

Each toy in the system has a unique \ToyId{}, which we call \texttt{ID} interchangeably in the rest of this paper. Over time, as the number of toys in the system grows, there is a risk of running out of IDs for new toys since the maximum \texttt{int32} is $2147483647$. To mitigate this risk, the type of \ToyId{} in the proto schema can be changed in-place from \texttt{int32} to \texttt{int64} shown in Listing \ref{lst:proto-schema-updated}.

\begin{lstlisting}[label={lst:proto-schema-updated}, float=ht, language=proto, numbers=none,
caption=Working example proto schema with \ToyId{} now \texttt{int64}.]
// A toy type known to our systems
message Toy {
  int64 toy_id; // **Changed to int64**
  uint32 price; // In US dollars
  bool in_inventory;
}
\end{lstlisting}

To be able to compile the code, this schema change first requires potential changes across the codebase everywhere \ToyId{} is directly or indirectly referenced, e.g. line 5 in Listing \ref{lst:working-code}. Additionally, for the code to properly handle \texttt{int64} values, line 13 needs to be updated to use $63$ instead of $31$. Finally, to ensure that the codebase properly works with int64 values instead of int32, it is a good idea to use large values (e.g. $123000000000$L) that exceed the maximum int32 value in our tests, i.e. modifications to lines 27, 36, 37 in Listing \ref{lst:working-code}.

\subsection{Manual Migration}
An instance of this migration for a specific ID has been undertaken at Google three years ago, where developers used regular expressions, \textit{regex} for short, such as \texttt{setToyId}, \texttt{getToyId}, \texttt{set\_toy\_id}, \texttt{get\_toy\_id} to find references to a specific ID field in a proto in the entire codebase for several programming languages including Java, C++ and Dart, and manually changed all the direct and indirect references to the ID across the entire codebase.

This effort took around two years to complete, and the changes have been carefully rolled out to production to avoid unexpected changes in any of the immediate and downstream systems.

Conducting this migration manually has been tedious, time consuming and stressful for the developers due to various factors.
\\

\noindent \textbf{Finding references accurately}: Using regex is a fairly simple technique to find references across the codebase. However, it suffers from several accuracy problems.

First, regex can match named variables, parameters and references throughout the code. However, it will likely find extra references that do not need changing, or will miss references that potentially need to be updated. As an example, line 13 in Listing \ref{lst:working-code} can be matched by a regex as \texttt{".*getToyId.*"}. When \ToyId{} is updated to \texttt{int64}, the existing code still compiles since \texttt{firstBitIsOne} on line 17 accepts a \texttt{long} value, creating unnecessary work for developers to manually investigate it. Furthermore, regex will not match the value $31$ on the same line, even though it needs to be updated to $63$.

Additionally, some ID names are quite short, and can match other unrelated IDs across the codebase. As an example, when migrating \ToyId{}, regular expressions might also match another irrelevant ID field named \texttt{toy\_id\_processed}. This makes manual investigation of referencing code even more tedious for developers.
\\

\noindent \textbf{Changing code manually}: Assuming developers carefully investigate and find all the code locations that need to be updated, they still need to check call sites and indirect references, and change all code manually, an error prone, tedious process. As an example, in Listing \ref{lst:working-code}, line 13 needs to be updated to use 63 instead of 31. Developers can miss such changes, especially when the number of locations to be updated is large.
\\

\noindent \textbf{Changing tests manually}: In addition to production code, there is typically accompanying test code that needs to be updated to catch potential problems in production code early. In Listing \ref{lst:working-code}, the ID of the test toy on lines 27 and 39 would benefit from updating to larger values. Such test code can easily be missed by developers if updated manually.
\\

\noindent \textbf{Verifying changes manually}: At Google, when developers make a code change, they typically run the regression test suite to verify that their changes did not inadvertently alter existing functionality. When regression tests fail, developers manually investigate the root cause of the failure. There are typically hundreds to thousands of changes across the codebase when migrating an ID, verifying these changes and investigating regression test failures at such scale seriously hinders developer productivity.
\section{Automated Migration}

\begin{figure*}
\centering
\includegraphics[scale=0.48]{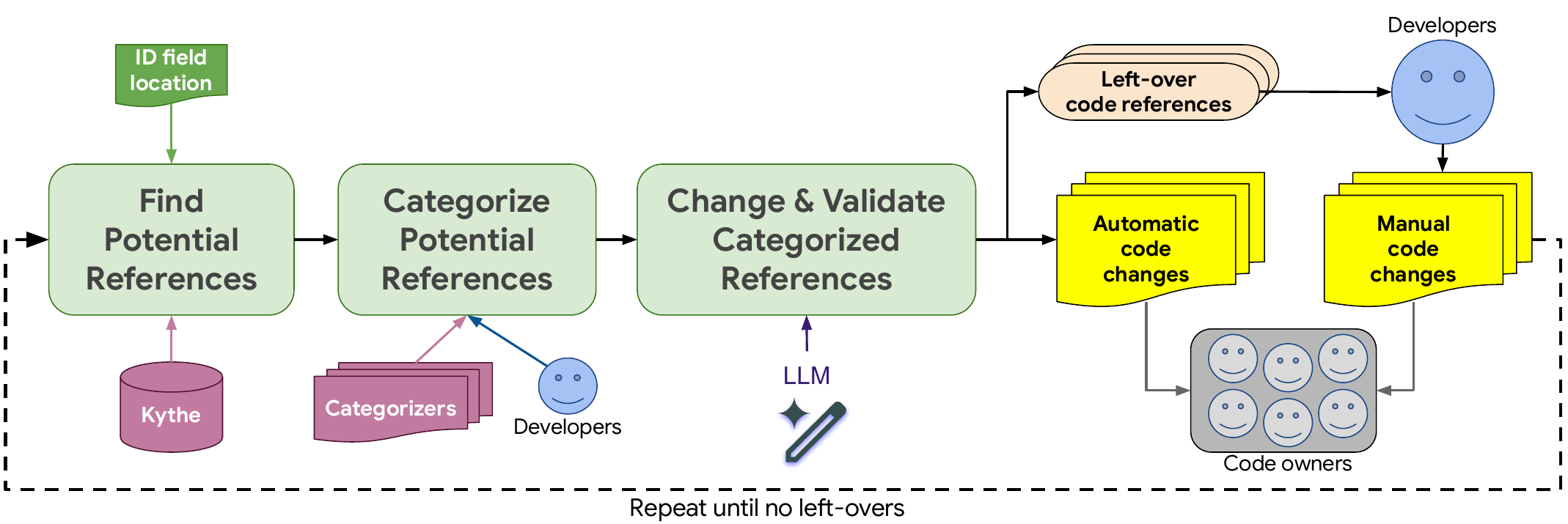}
\caption{High level overview of the automated ID migration system. The system runs nightly, finds potential ID references and categorizes them. Developers make the necessary code changes using an LLM or manually (if the LLM failed to make the change), and send the changes to code owners. This process continues until no more ID references are left over to migrate.}
\label{fig:system-overview}
\end{figure*}

Due to the difficulties in performing manual migrations discussed in the previous section, we built a system that leverages several components to perform ID migrations mostly automatically end to end, summarized in the high level overview in Figure \ref{fig:system-overview}. We previously shared high-level information about this work \cite{ai-migrations-blog-post,ai-migrations-icse-2025}, in this paper we discuss our approach, success metrics, results and learnings in great detail.

This system runs nightly until all locations are processed and there are no outstanding locations left to be migrated. A migration does not follow a waterfall model to be fully completed. Instead it is typically a continuous process where code locations are migrated over time over many sessions potentially by several developers until there are no more locations to migrate.

We discuss each component of the system in the sections below.

\subsection{Find Potential References}

The initial input for a migration is an ID field, similar to \ToyId{} discussed in Listing \ref{lst:proto-schema}. Google uses a monolithic code repository \cite{monorepo, monorepo-2}, and a code indexing system called Kythe \cite{Kythe} that indexes the entire codebase in the repository. Using Kythe, we first find the direct references to the ID, then find the direct references to those references, and keep finding other indirect references to the ID in the entire codebase for a maximum total distance of five. There are some important implications of this approach.
\\

\noindent \textbf{Irrelevance}: Our approach finds direct and indirect ID references, some of which actually do not need to be changed. As an example, in Listing \ref{lst:working-code}, our approach finds lines 5, 13, 27, 39 as direct references, lines 2, 12, 25, 35, 37 as distance two indirect references since they contain the direct references, and lines 1, 22, 36, 42 as distance three indirect references as they either contain or call the distance two references.

This list contains code locations that do not necessarily need to be modified during the migration, i.e. we opt to be conservative to avoid missing any references and use a super-set of the locations that require changing.

We tackle this challenge using several techniques, e.g. classifying locations with automation, and using an LLM to output its decision on identified locations, discussed in Section \ref{section:categorize-potential-references}.
\\

\noindent \textbf{Incompleteness}: After a certain distance, the list of references grows quite large with little upside regarding the relevance of the found locations. Therefore, due to practical reasons, our approach considers ID references up to a distance of five, beyond which it can potentially miss some references.

We tackle this challenge by running regression test suites of the code owning teams, discussed in Section \ref{section:update-code}, and in some cases, by rolling out changes in production slowly to observe any potential adverse effects, discussed in Section \ref{section:discussion}.
\\

\noindent \textbf{Precision}: Direct references to an ID do not always constitute the exact locations in code to be modified. In Listing \ref{lst:working-code}, line 39 is identified as a direct reference to \ToyId{}, although the actual changes need to be performed on lines 36 and 37, as they contribute to the reference on line 39 through data flow. Our approach is not precise in exactly identifying lines 36 and 37, but precise enough to identify line 39, close to the actual needed change locations.

We tackle this challenge by relying on an LLM to consider the files to be modified in their entirety and to make changes anywhere in the passed file, even though we are not precise on the exact lines to be changed in our prompt to the LLM, discussed in Section \ref{section:update-code}.
\\

It is possible to overcome these accuracy problems in the identified potential references using approaches such as data flow analysis \cite{kildall1973unified} and using an AST \cite{ast} parser. However, these techniques are potentially computationally expensive, and require implementing and maintaining logic to predict and cover many potential code variations written in different styles by thousands of developers over many years, a costly, imprecise and practically intractable approach. As a result, we opted to take an inaccurate yet practical approach to finding potential references to the ID, and using an LLM to bridge the gaps in accuracy, discussed in the later sections.

\subsection{Categorize Potential References}
\label{section:categorize-potential-references}
After finding the potential references, we categorize them into buckets based on confidence on whether they need to be migrated or not. The categories are listed below with example categorizers used to categorize each location.
\\

\noindent \textbf{Not-migrated}: This is a location that is identified to have not been migrated with 100\% confidence. A typical categorizer that identifies a test location as not-migrated uses a regex to match the value used in accessor methods. As an example, in Listing \ref{lst:working-code} line 27, a regex checks the value $-5$, identifies it as a value inside the int32 range in Java, and marks it as not-migrated. There are more categorizers for both production and test code, and for different programming languages.
\\

\noindent \textbf{Irrelevant}: This is a location that is determined to be irrelevant with 100\% confidence. As an example, a regex categorizer would mark line 27 in Listing \ref{lst:working-code} as irrelevant, if the value were to be outside the int32 range in Java. Another categorizer  categorizes lines 1, 2, 12, 22, 25, 35 as irrelevant, as these are either class or method definitions, and are not relevant to the migration. Finally, developers investigate "left-over" locations discussed below, and may decide that specific locations do not require migration by marking them as irrelevant explicitly in the configuration of the system.
\\

\noindent \textbf{Relevant}: This is a location that is not marked as "not-migrated" or "irrelevant", yet it is a reference to the ID field, and needs to be investigated. In Listing \ref{lst:working-code}, lines 13 and 39 are marked as relevant, since they cannot be marked as not-migrated or irrelevant by earlier categorizers, yet they access \ToyId{}, and should potentially be investigated further.
\\

\noindent \textbf{Left-over}: This is a location that is not categorized into any of the previous categories and is left over to be investigated by developers manually. In their investigation, developers may decide whether a location needs to be migrated or is irrelevant, which they dictate through configuration. In the next run of the system, the investigated location is moved into the dictated category, not-migrated or irrelevant.

\subsection{Change \& Validate Categorized References}
\label{section:update-code}
To make the code changes, we leverage an internal version of the Gemini \cite{team2023gemini} model fine-tuned on internal Google code and data \cite{monorepo, monorepo-2} using the DIDACT framework \cite{didact} with multiple tasks for software engineering, including code-review comment resolution, edit prediction, variable renaming, code-review comment prediction, and build-error repair \cite{frommgen2024resolving}. Although the LLM model has continuously been updated, we used the same fixed version of it with \texttt{temperature=0.0} throughout the course of our study for stability, we did not fine-tune the model for our specific task, and only used prompting for the code modifications.

After categorizing potential references into buckets, we use the LLM to make code changes on the locations categorized as Not-migrated and Relevant, by using language-appropriate prompts, shown in Listing \ref{lst:prompt-java} for Java, and passing the entire file to be modified with the suggested lines of the respective locations to the LLM.

The full contents of the file \texttt{ToyView.java} in Listing \ref{lst:working-code} and line numbers 5, 13 is passed to the LLM for modifications. Similarly, the full contents of the file \texttt{ToyViewTest.java} and line numbers 27, 36, 37, 39, 42 are passed to the LLM constructing the model prompt. The passed line numbers are suggestive, and the LLM has flexibility on making changes in other lines anywhere in the file. 
The line suggestions are represented in the model prompt as comments instructing the model of what should be changed in the given line.

\begin{figure}
\centering
\includegraphics[scale=0.42]{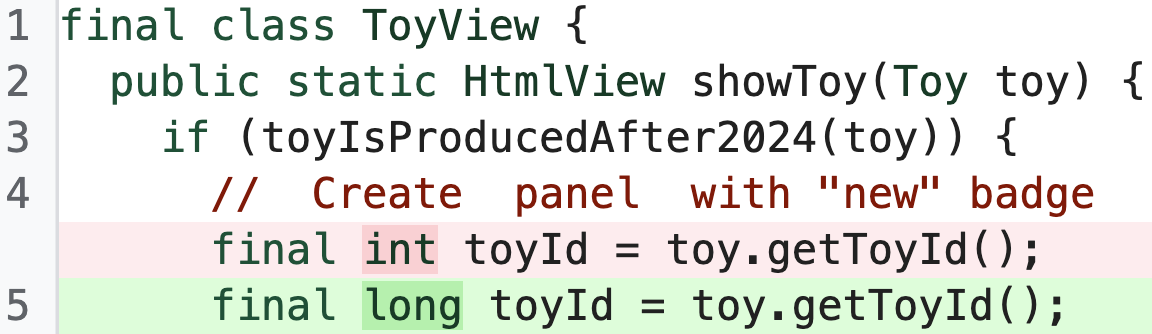}
\caption{Example edit on a file in a code change and its diff in the UI of Critique \cite{critique}, the web-based code review system used at Google. Deleted code is in red, added code is in green, \Leven{} \cite{levenshtein} edit distance (\DELTA{}) is 4.}
\label{fig:code-diff}
\end{figure}

The LLM outputs changes in the form of \textit{diffs} \cite{diff, hunt1976algorithm}, which are applied to the original file using a fuzzy diffing algorithm. This finds the locations of code to replace by using edit distance to minimize simple LLM errors.
An example diff is shown in Figure \ref{fig:code-diff} in the UI of Critique \cite{critique}, the web-based code review system used at Google, where deleted code is in red, and added code is in green. 

Since LLMs are non-deterministic, three rounds of modifications are attempted for each file, and a random change from among the successful attempts is chosen for each file.

\begin{lstlisting}[label={lst:prompt-java}, float, breaklines=true, breakindent=0pt, frame=single, language=text, numbers=none, caption=Prompt to update \ToyId{} references in production and test code in Java.]
toy_id should be of type long. Update the code and respective references properly.

Also update the tests to reflect a large id. Initialize the toy_ids with values larger than 10000000000, e.g. if the previous id was 1, it should now be 10000000001L. If the previous id was negative, the new value should also be negative.
\end{lstlisting}

As LLMs may produce erroneous code \cite{dou2024s, wang2024large}, the LLM's response and its changes on an entire file are validated using several strategies in the following order:
\begin{enumerate}
    \item \textbf{Success}: If the LLM did not return a successful completion, the change is invalid. There are several reasons that may cause an unsuccessful completion response: the LLM server may be down, the calling system's LLM resource quota may be exhausted, the size of the input file may exceed the LLM's allowed context length limit.
    \item \textbf{Whitespace}: If the LLM only changed whitespaces, i.e. it simply reformatted the code, the change is invalid.
    \item \textbf{AST parser}: If the changed file cannot be parsed with an AST parser, i.e. it is malformed, or the AST of the changed file is identical with the original, the change is invalid.
    \item \textbf{Punt}: We pass the files' contents before and after the change, and ask the LLM whether the change was necessary. If the LLM punts by responding "No", the change is invalid. 
    \item \textbf{Build}: If the changed file causes build breakages during compilation, the change is invalid.
    \item \textbf{Test}: If the changed file causes regression test failures, the change is invalid.
\end{enumerate}

The validations are conducted in the listed order from less expensive to more expensive. If the changed file fails any of the validations, it is discarded without running further validations, and marked to be migrated manually by developers. If the file is successfully modified by the LLM and passes all validation steps, it is marked to be ready for review.

In the end, all changed files, automatically modified by the LLM or manually modified by developers, are grouped into code changes with potentially several changed files per change, each change is visually investigated by developers for verification, any necessary final changes are manually performed within the code change by developers, and finally each code change is sent to the owners of the code for final acceptance and submission through Critique \cite{critique}, shown in Figure \ref{fig:code-diff}.

\section{Evaluation and Discussion}

In this section, we discuss how we evaluate our system, describe our case studies, and report the results of our interviews on the usability of our system.

\subsection{Evaluation Metrics: \LlmDelta{} and \HumanDelta{}}

To evaluate the effectiveness of the automation of our system, we use the number of characters edited by humans and LLMs to complete an ID migration. The output of our system is a code change, with one or more changed files in it. A code change falls into one of three categories:
\\

\noindent \textbf{LLM-Only}: The code change only contains files entirely modified by the LLM.
\\

\noindent \textbf{LLM-then-Human}: The code change is initially generated with the files modified by the LLM, but humans changed the files further upon visual investigation of the changes.
\\

\noindent \textbf{Human-Only}: The code change only contains files entirely modified by developers manually, with no involvement from the LLM.
\\

\begin{figure}
\centering
\includegraphics[scale=0.45]{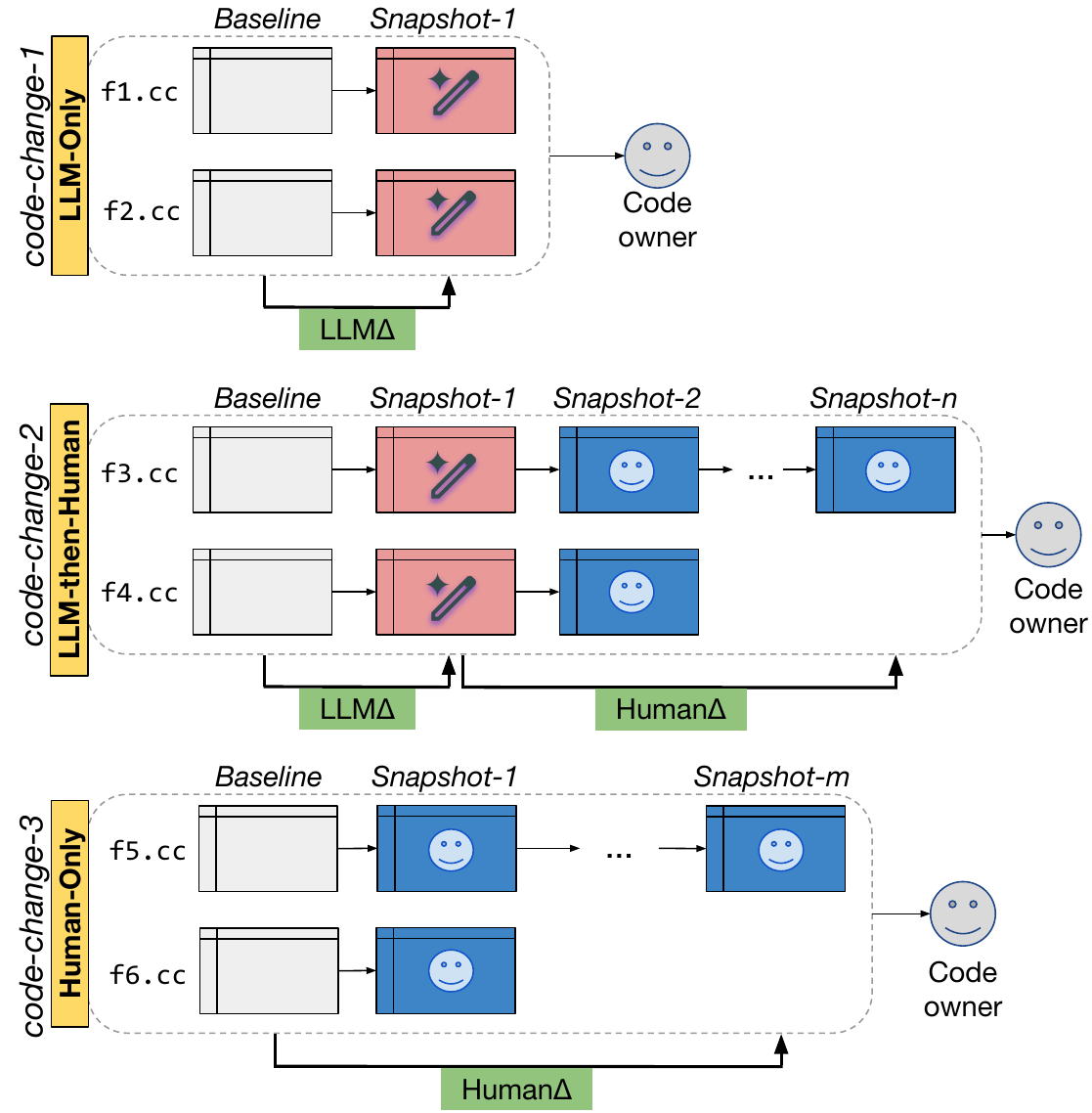}
\caption{Types of code change, and how the number of edited characters by the LLM and developers are calculated.}
\label{fig:code-change-types}
\end{figure}

Examples of each code change type are shown in Figure \ref{fig:code-change-types}, where \textit{code-change-1} is LLM-Only, \textit{code-change-2} is LLM-then-Human, and \textit{code-change-3} is Human-Only.

In all code changes, \textit{Baseline} is the version of a file that is not yet changed. Once a file is modified and saved, a \textit{snapshot} is created. Subsequent further modification and save events result in new snapshots. Between each snapshot, there are character edits to change the code to the new snapshot, shown in Figure \ref{fig:code-diff}, where we use \Leven{} \cite{levenshtein} edit distance, \DELTA{} for short, as a proxy to measure the effort required to make the modifications.

In \textit{code-change-1}, \textit{Snapshot-1} contains the automated LLM changes. The \Leven{} edit distance between \textit{Baseline} and \textit{Snapshot-1} is calculated as the \LlmDelta{}.

In \textit{code-change-2}, developers make further edits in each file in addition to the LLM changes, so the subsequent snapshots from \textit{Snapshot-2} to the final snapshot submitted with the code change contain human edits. We calculate the \Leven{} edit distance between \textit{Baseline} and \textit{Snapshot-1} as the \LlmDelta{}, and the \Leven{} edit distance between \textit{Snapshot-1} and \textit{Snapshot-n} as the \HumanDelta{}. Note that, the \HumanDelta{} is not an aggregation of all the edits between each snapshot, but instead the single edit distance between \textit{Snapshot-1} and \textit{Snapshot-n}, as the edits between the intermediate snapshots are considered work in progress, and are ignored for a fair comparison between \LlmDelta{} and \HumanDelta{}.

In \textit{code-change-3}, all snapshots from \textit{Snapshot-1} to \textit{Snapshot-m} contain human edits, as the LLM is not involved in the edits for Human-Only code changes at all. As a result, the \Leven{} edit distance between \textit{Baseline} and \textit{Snapshot-m} is calculated as the \HumanDelta{}.

We track every code change in both categories, calculate the total edits across all files in all code changes related to the migration of an ID, and compare the \LlmDelta{} and \HumanDelta{} for the entire ID.

\subsection{Statistics On Migrated IDs}

Summarized in Table \ref{table:migration-info}, \TotalClsDevsText{} developers migrated \TotalIds{} distinct IDs over \TotalMonths{} months using our system, a total of \TotalCls{} code changes have been submitted, \TotalClsLlmPct{} of the code changes (LLM-Only and LLM-then-Human) have been generated by the LLM, a total of \TotalEdits{} edits across all code changes have been performed, and \TotalEditsLlmPct{} of those edits were by the LLM.

\begin{table}
\centering
\caption{Aggregate statistics about the migrations.}
\begin{tabulary}{\linewidth}{R|L}
\# IDs migrated & \TotalIds{} \\
\# Developers that conducted migrations & \TotalClsDevs{} \\
\hline
\underline{Total code changes} & \TotalCls{} \\
by developer-1 & \TotalClsDevOne{} \\
by developer-2 & \TotalClsDevTwo{} \\
by developer-3 & \TotalClsDevThree{} \\
\hline
\underline{Total code changes} & \TotalCls{} \\
LLM-Only & \TotalClsLlmOnly{} (\TotalClsLlmOnlyPct{}) \\
LLM-then-Human & \TotalClsLlmThenHuman{} (\TotalClsLlmThenHumanPct{}) \\
Human-Only & \TotalClsHumanOnly{} (\TotalClsHumanOnlyPct{}) \\
\hline
\underline{Total code changes} & \TotalCls{} \\
\# reviewers & \TotalClReviewers{} \\
\# teams & \TotalClTeams{} \\
\# offices & \TotalClOffices{} \\
\# time zones & \TotalClTimezones{} \\
\hline
\underline{\TotalDelta{} across all IDs} & \TotalEdits{} \\
\LlmDelta{} & \TotalEditsLlm{} (\TotalEditsLlmPct{}) \\
\HumanDelta{} & \TotalEditsHuman{} (\TotalEditsHumanPct{}) \\
\end{tabulary}
\label{table:migration-info}
\end{table}

\begin{table}
\centering
\caption{Statistics for each migrated ID.}
\begin{tabulary}{\linewidth}{p{0.75cm}|>{\raggedleft\arraybackslash}p{1.4cm}|C|C|C}
   & \TotalDelta{}$\downarrow$ & \# Code \newline Changes & \# Modified \newline Files & \# Programming \newline Languages \\ \hline
ID-30 &  23800    &     39      &      418      &       2         \\
ID-32 &  22822    &     185     &      392      &       5         \\
ID-20 &  16406    &     28      &      175      &       2         \\
ID-28 &   7059    &     29      &      89       &       1         \\
ID-03 &   4594    &     21      &      350      &       4         \\
ID-37 &   2774    &     25      &      58       &       1         \\
ID-06 &   2458    &     23      &      41       &       2         \\
ID-35 &   2039    &     14      &      38       &       2         \\
ID-19 &   1937    &     17      &      21       &       1         \\
ID-18 &   1889    &     20      &      33       &       1         \\
ID-29 &   1825    &     6       &      38       &       1         \\
ID-04 &   1279    &     10      &      27       &       1         \\
ID-31 &   1223    &     15      &      31       &       1         \\
ID-26 &   1151    &     4       &      11       &       1         \\
ID-34 &   1088    &     4       &      16       &       2         \\
ID-09 &   1038    &     6       &      16       &       1         \\
ID-38 &   844     &     8       &      17       &       2         \\
ID-14 &   802     &     7       &      12       &       2         \\
ID-01 &   798     &     16      &      103      &       4         \\
ID-07 &   673     &     7       &      31       &       1         \\
ID-05 &   584     &     5       &      9        &       1         \\
ID-08 &   578     &     6       &      13       &       2         \\
ID-22 &   571     &     2       &      19       &       1         \\
ID-17 &   538     &     4       &      6        &       1         \\
ID-36 &   386     &     5       &      8        &       2         \\
ID-12 &   374     &     5       &      40       &       4         \\
ID-02 &   367     &     10      &      14       &       1         \\
ID-23 &   326     &     10      &      15       &       1         \\
ID-33 &   299     &     7       &      7        &       2         \\
ID-13 &   270     &     8       &      14       &       2         \\
ID-27 &   221     &     5       &      6        &       1         \\
ID-15 &   221     &     11      &      12       &       2         \\
ID-16 &   156     &     9       &      13       &       2         \\
ID-21 &   146     &     3       &      10       &       1         \\
ID-25 &   128     &     7       &      11       &       1         \\
ID-10 &   98      &     1       &      4        &       1         \\
ID-11 &   95      &     7       &      7        &       2         \\
ID-24 &   26      &     2       &      2        &       1         \\
ID-39 &   23      &     1       &      1        &       1         \\
\end{tabulary}
\label{table:migration-info-by-id}
\end{table}

IDs have varying characteristics regarding the programming languages that reference the ID and the migration size, listed in Table \ref{table:migration-info-by-id}, where IDs are sorted by descending \TotalDelta{}. While some IDs required edits spread across many files (e.g. ID-30, ID-32, ID-03), some IDs were more localized into a few files (e.g. ID-10, ID-24, ID-27).

\begin{figure}
\centering
\includegraphics[scale=0.36]{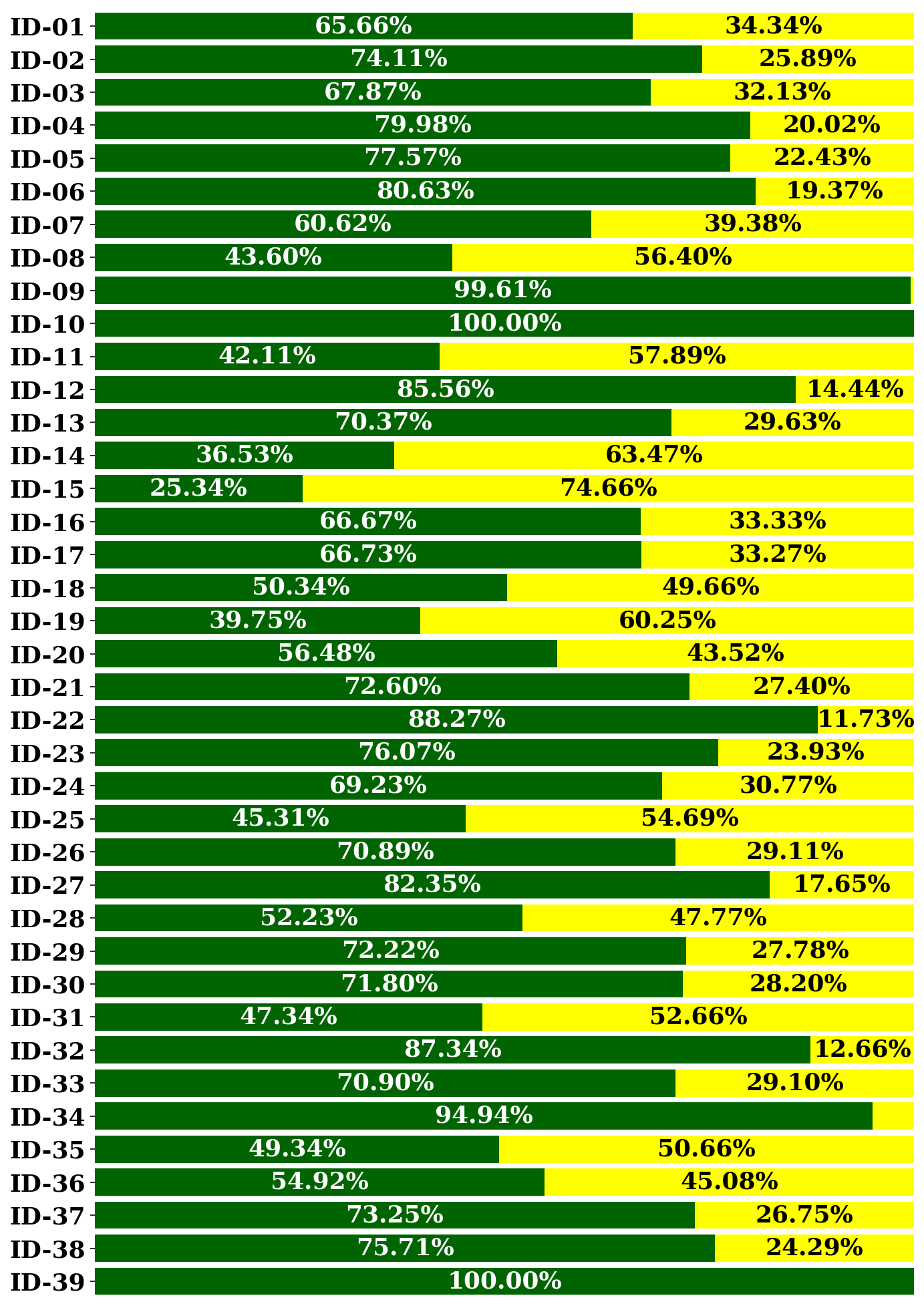}
\caption{\LlmDelta{} and \HumanDelta{} percentages for each ID.}
\label{fig:llm-vs-human}
\end{figure}

Additionally, \LlmDelta{} and \HumanDelta{} split is summarized as a percentage of total edits for each ID in Figure \ref{fig:llm-vs-human}.

\subsection{Discussion}
\label{section:discussion}

At the end of the migrations, we collected the statistics discussed in the previous section and conducted interviews with the \TotalClsDevsText{} developers that performed the migrations using our system. In this section, we summarize and discuss the developers' observations and experiences, the advantages and disadvantages they reported using the system, and highlight important differences in the characteristics of the migrated IDs.
\\

\noindent \textbf{Automation benefits}: The developers reported great satisfaction with the end-to-end automation of the code changes, especially when the LLM worked well and suggested the correct modifications. Our system ran nightly, and promoted the code changes that passed all validations to be reviewed by developers. Then, the developers investigated the code changes for final verification and sent them out to code owners. \TotalClsLlmPct{} of the code changes have been automatically created by the system, resulting in high productivity.

For changes that they had to manually create, the developers still reported high satisfaction with the accuracy of the potential references found for them to investigate, compared to their previous regex based approach. They estimated that, compared to their previous attempts, this system reduced the total time spent on a migration by \TotalTimeReduction{}. We do not have concrete wall-clock effort measurements on the total time spent, although the percentage of automatically created code changes, \TotalClsLlmPct{}, and the total \LlmDelta{} across all migrations, \TotalEditsLlmPct{}, directionally align with the developers' estimate.

Additionally, as our system runs nightly and updates the remaining left-over references automatically using the submitted code changes, the developers reported to have a continuously renewed state of the migration, providing a sense of progress towards the completion of the migration.
\\

\noindent \textbf{Validation benefits}: The developers highlighted automated validations as a significant improvement on their productivity. Before automation, in addition to their manual edits, developers needed to create the code change, start validations manually, and wait for them to finish before they decide whether the code change is ready for review by code owners. Our system only promotes those code changes that pass all validation steps that are ready for review, decreasing the tedious work developers need to undertake.
\\

\noindent \textbf{LLM benefits}: The developers reported high satisfaction with the use of an LLM to suggest modifications. There are a myriad different ways developers write code, and it is practically prohibitive to predict all potential uses and expensive to declare and maintain code transformations with popular rule based tools such as TXL \cite{cordy2004txl}, Stratego/XT \cite{bravenboer2008stratego}, Rascal \cite{klint2009rascal}.

\begin{figure}[tbp]
    \centering
    
    \begin{minipage}{\linewidth}
        \centering
        \includegraphics[width=\linewidth]{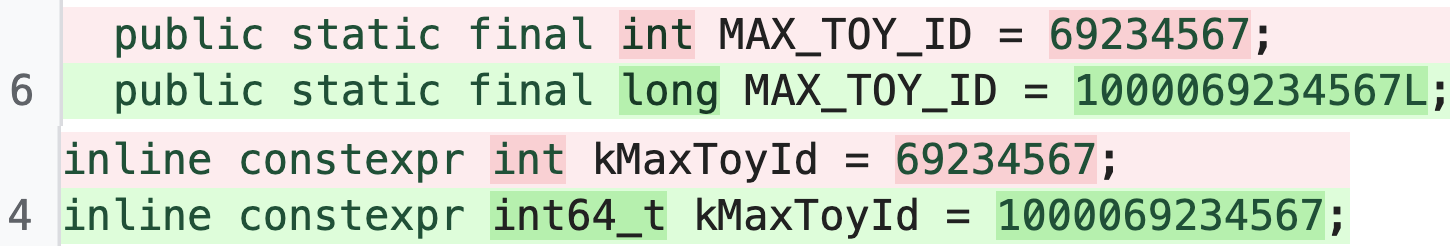}
        \subcaption{Changes in multiple languages with almost identical prompt}\label{fig:llm-change-examples-a}
    \end{minipage}
    
    \vspace{0.3cm}
    
    \begin{minipage}{\linewidth}
        \centering
        \includegraphics[width=\linewidth]{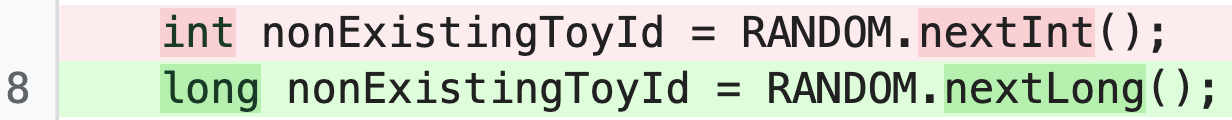}
        \subcaption{Language specific domain knowledge}\label{fig:llm-change-examples-b}
    \end{minipage}
    
    \vspace{0.3cm}
    
    \begin{minipage}{\linewidth}
        \centering
        \includegraphics[width=\linewidth]{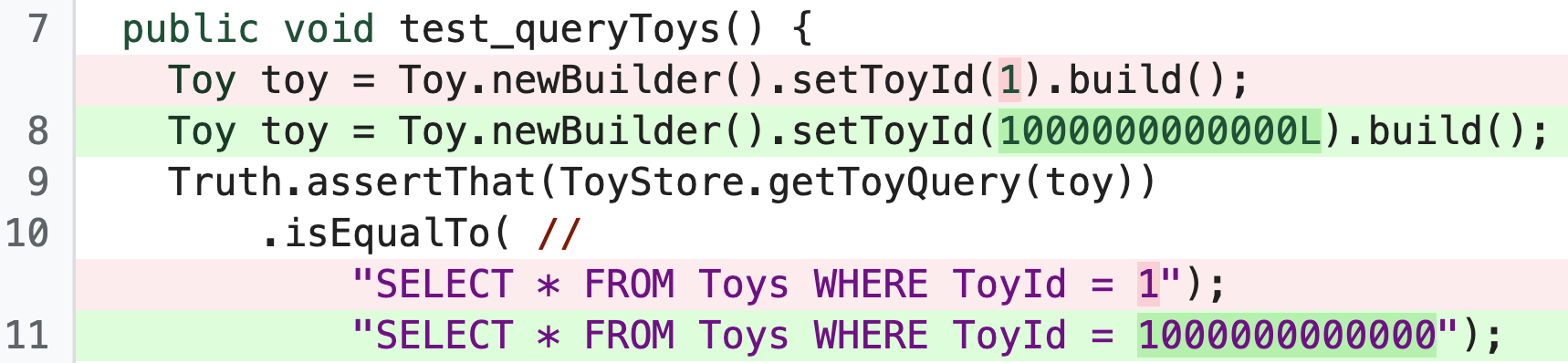}
        \subcaption{Updating values in code and in SQL strings together}\label{fig:llm-change-examples-c}
    \end{minipage}
    
    \vspace{0.3cm}
    
    \begin{minipage}{\linewidth}
        \centering
        \includegraphics[width=\linewidth]{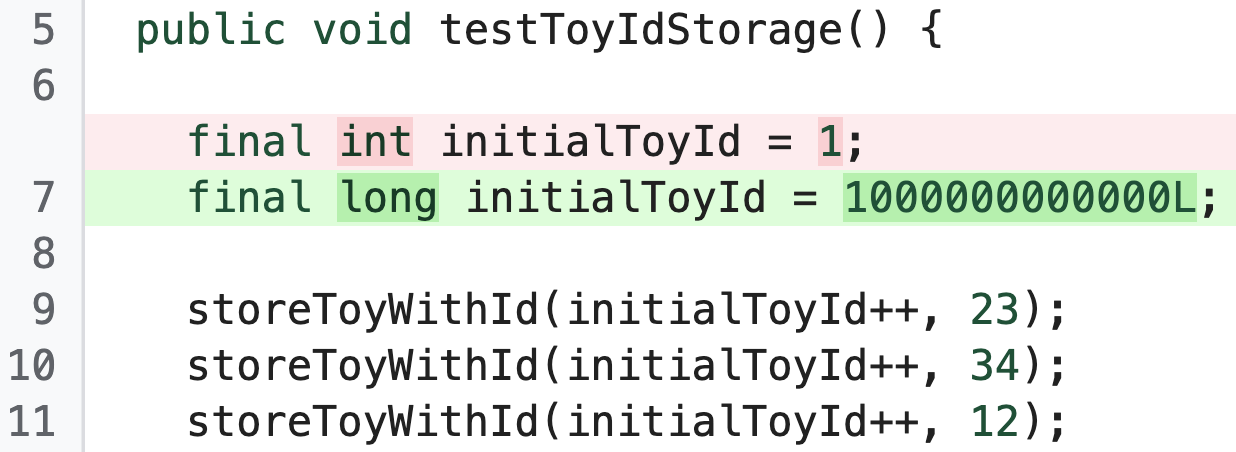}
        \subcaption{Making the minimal change instead of changing all lines}\label{fig:llm-change-examples-d}
    \end{minipage}
    
    \caption{Examples of LLM edits.}
    \label{fig:llm-change-examples}
\end{figure}

LLMs are reported to have good representation and reasoning on code \cite{fried2022incoder,li2023starcoder,nijkamp2022codegen,achiam2023gpt}, making it possible to ask for intended changes using a standard prompt in English to tackle a wide variety of code patterns. Several examples where production code is adapted to our working example to mirror actual submitted code changes, are shown in Figure \ref{fig:llm-change-examples}, demonstrating the LLM's domain knowledge, flexibility, and its ability to change relevant code.

Figure \ref{fig:llm-change-examples-a} shows an example test where an almost identical prompt results in the successful update of Java and C++ code. In Figure \ref{fig:llm-change-examples-b}, the LLM exhibits domain knowledge on the appropriate method to call in Java after the migration. Figure \ref{fig:llm-change-examples-c} shows an example where the LLM correctly identifies and changes not only code but also its counterpart in the SQL string. Finally, in Figure \ref{fig:llm-change-examples-d}, even though the LLM is passed three distinct line numbers to change in the file, lines 9, 10 and 11, and even though those lines contain multiple numbers that could potentially be confusing, the LLM identifies the right number to change, i.e. \texttt{initialToyId}, and makes the minimal change of updating line 7, instead of the three lines suggested.

Developers reported satisfaction on the LLM's ability to flexibly cover such different cases in different programming languages, removing the burden of implementing and maintaining code transformation rules or tooling.
\\

\noindent \textbf{LLM non-determinism challenges}: LLMs exhibit non-deterministic behavior, resulting in varied output diffs even when presented with identical inputs across multiple executions. This inherent variability elicited mixed reactions from developers, expressing satisfaction and dissatisfaction under different use-cases. When the LLM does not work on a file, developers reported satisfaction on the ability to re-run it on the same inputs several times and to sometimes get a satisfactory code change in the end. This unpredictability also led to dissatisfaction when the LLM failed, as developers could not anticipate how many retries would be needed before resorting to manual changes, and they were unhappy about any manual work they had to carry out in the end.
\\

\noindent \textbf{LLM context-window challenges}: One of the limitations of using an LLM is the input context window size, some edits were not performed by the LLM as the entire contents of a large file did not fit into the context window\footnote{This is mitigated in the latest Gemini versions with a much larger context window}. Such change attempts failed in the first validation step "Success" in Section \ref{section:update-code}, and those changes had to be performed by the developers manually.
\\

\begin{figure}[tbp]
    \centering
    
    \begin{minipage}{\linewidth}
        \centering
        \includegraphics[width=\linewidth]{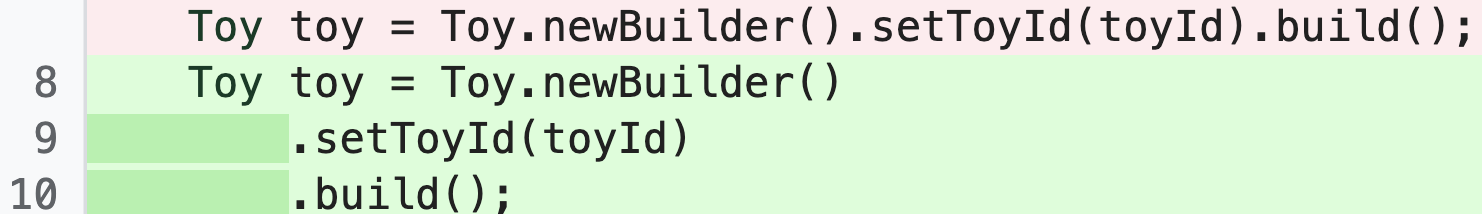}
        \subcaption{Hallucination that reformats file contents}\label{fig:llm-hallucination-examples-a}
    \end{minipage}
    
    \vspace{0.3cm}
    
    \begin{minipage}{\linewidth}
        \centering
        \includegraphics[width=\linewidth]{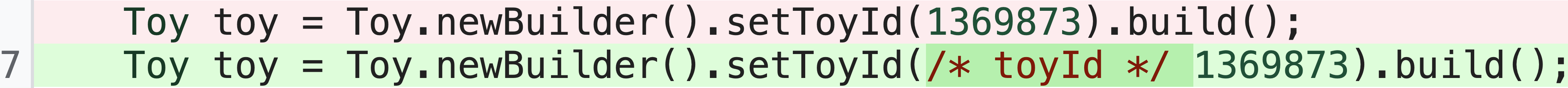}
        \subcaption{Hallucination that adds comments}\label{fig:llm-hallucination-examples-b}
    \end{minipage}
    
    \caption{Examples of LLM hallucinations.}
    \label{fig:llm-hallucination-examples}
\end{figure}

\noindent \textbf{LLM hallucination challenges}: The LLM sometimes had hallucinations, where it simply reformatted the code as in Figure \ref{fig:llm-hallucination-examples-a}, only added comments to the code as in Figure \ref{fig:llm-hallucination-examples-b}, or made entirely irrelevant changes. Developers had to manually handle such cases, constituting \TotalClsHumanOnlyPct{} of the code changes and \TotalEditsHumanPct{} of the edits.
\\

\noindent \textbf{LLM language support challenges}: Certain languages, including Java, C++, and Python were well supported by the used LLM, while Dart was not well supported, as the LLM was not trained on enough Dart code. As a result, IDs with heavy Dart use, e.g. ID-15, ID-31 and ID-35, ended up with lower \LlmDelta{} percentages.
\\

\noindent \textbf{Ramp-up}: When the developers first started the migrations, they were not familiar with using our system. To maximize the benefits of automation provided by the system, they started with smaller, localized IDs. Until they familiarized themselves to use the system effectively, they ended up making some of the edits manually. As they gained more experience, they moved on to larger, more spread IDs, with higher use of automation. This directly contributed to the differences between the \LlmDelta{} and \HumanDelta{} split observed across different IDs shown in Figure \ref{fig:llm-vs-human}. For instance, ID-24 and ID-25 have low \TotalDelta{}, but also low \LlmDelta{} percentages, while ID-30 has the highest \TotalDelta{}, yet it has a high \LlmDelta{} of 71.80\%.
\\

\noindent \textbf{Pre-existing failures}: At Google, each code change goes through regression testing before submission. If the regression test suite has pre-existing failures unrelated to the code change under review, validation of the code change fails at the "Test" step discussed in Section \ref{section:update-code}. This has been a hindrance for the developers during migrations, even though the LLM has been successful in some of the code changes, those code changes were not promoted for review, and developers had to contact the code owners to fix the pre-existing failures and only after then made the code changes related to the ID migration.
\\

\noindent \textbf{Pre-existing goldens}: At Google, expected outputs of tests, called \textit{goldens}, are sometimes stored in the source code repository along with characterization tests \cite{feathers2004working}. When test code is updated, goldens typically require manual updates from developers by running specialized tools. As our automated system did not support running such tools, tests using goldens sometimes failed the "Test" step discussed in Section \ref{section:update-code}, and required manual work from developers, contributing to lower \LlmDelta{} percentages.
\\

\noindent \textbf{Production roll-outs}: Small, localized ID migrations were typically completed by the developers entirely without a need for additional work . However, large, distributed IDs, a total of 2 out of the \TotalIds{} migrated, namely ID-30 and ID-32, required attention from the code owners before they were rolled out into production. As the impact of such IDs potentially spanned across several teams, changes were rolled out slowly to observe any adverse effects. Our system provided no automation or benefits for this step.
\\

The findings of this study, encompassing both the advantages and disadvantages of our system, alongside developer feedback, indicate that our approach strikes an effective balance between automation and user oversight for the specific migration task within Google's infrastructure. We posit that the high-level workflow implemented in our system, especially using an LLM but verifying its proposed changes through tooling, could be adapted for various types of migrations in different development environments, thereby providing a foundation for LLM-assisted automation to support developers.

\section{Threats To Validity}
In this section, we discuss the threats to the validity of our work and case studies.
\\

\noindent \textbf{Case study}: Our case study does not involve a controlled experiment comparing the LLM-based approach to other techniques, making it difficult to isolate the specific contribution of the different components of the system to the observed improvements.

Additionally, our case study was conducted within the specific context of Google's development environment and infrastructure, which may limit the generalizability of the findings to other organizations.

Our case study involved \TotalIds{} ID migrations conducted by \TotalClsDevsText{} developers, which may not be representative of the full range of other types of code migration tasks and developer experiences.

Finally, our case study focuses on a specific type of code migration, migrating identifiers from 32-bits to 64-bits. Our results may not generalize to other types of migrations.
\\

\noindent \textbf{Developer Bias}: The developers involved in the case study were also involved in the development of the system by providing feedback, potentially introducing bias in their assessment of the system's effectiveness.
\\

\noindent \textbf{LLM}: Our study uses an LLM developed to be used internally at Google. Although popular large LLMs have been shown to perform well on tasks such as summarization and question answering, the generalizability of our results on migrations to other LLMs may be limited.
\\

\noindent \textbf{Effort estimation}: The study uses \Leven{} edit distance as a measure of developer effort, which may not fully capture the cognitive load and complexity involved in code modifications.

Furthermore, the reported \TotalTimeReduction{} time savings is based on developer perception and estimations, and not on precise time tracking data. Although their estimations are in line with the proportion of automatically generated code changes and edits, there may be potential inaccuracies in their assessment of efficiency gains.

\section{Related Work}

There has been extensive research in literature on manipulating code under different domains including code migration, code transformation, and code refactoring, all relevant to our work. Code translation is, to a degree, another relevant domain to the work discussed in this paper.
\\

\noindent \textbf{Code translation}: Early research on code translation relied on manually defined rules and various program analysis techniques that operate on different abstractions on the source code.

There has been work on using parse trees \cite{chiswell2007mathematical} and abstract syntax trees \cite{ast} generated from source code, transformed and fed to the target compiler \cite{feldman1990fortran, sharpen, c2rust, JavaToCSharp}. Yasumatsu and Doi \cite{yasumatsu1995spice} built a system that creates runtime replacement classes between Smalltalk and C to implement the same functionality. Baker \cite{baker1996parameterized} uses parameterized pattern matching techniques on a stream-of-strings representation of source code to map patterns to the destination language. SCRUPLE \cite{paul1994framework} uses regular-expressions to locate and translate programming patterns between languages. Johnson \cite{johnson1994substring} uses hash-based fingerprints to identify text-level patterns in source code to be translated to the target language. Syntax based editing approaches have also been used for code translation \cite{chen1990c, steffen1985interactive, ballance1992pan, borras1988centaur, reps1984synthesizer, ladd1995language}. These approaches require either manually creating rules for translation or implementing specific tooling, a time consuming task with potentially unintended results including low readability and low accuracy of the translated code \cite{kontogiannis2010code}.

More recently, effectiveness of LLMs for code translation tasks has been investigated extensively. Weixiang et al. \cite{yan2023codetransocean} propose a benchmark to assess the success of code translation tasks. SteloCoder \cite{pan2023stelocoder} uses LLMs to translate code into Python, while FLOURINE \cite{eniser2024towards} translates source code to Rust. TRANSAGENT \cite{yuan2024transagent} and TransCoder \cite{roziere2020unsupervised} translate code between several programming languages using automated validation and fixing mechanisms to improve accuracy. UniTrans \cite{yang2024exploring} uses test-cases as a benchmark and to guide code translation between multiple programming languages. CoTran \cite{jana2023cotran} augments LLMs to translate code by using symbolic execution based testing and reinforcement learning on the LLM outputs. A recent study by Pan et al. \cite{pan2024lost} points out the shortcomings of LLM-based code translation approaches by demonstrating various mistakes LLMs make.

Our work is different from code translation, as we modify code within the same language to obtain minor differences in its behavior.
\\

\noindent \textbf{Code transformation}: Early research on code transformation has been rule-based where transformations are specified using manually defined rules, such as TXL \cite{cordy2004txl}, Stratego/XT \cite{bravenboer2008stratego}, Rascal \cite{klint2009rascal}.

Gabel and Su \cite{gabel2010study} showed high redundancy and repetitiveness within short windows of source code when analyzed with syntactical tokens. This led to further studies \cite{nguyen2013study} that exploit the repetitiveness to suggest transformation by example \cite{lieberman2001your}, where an example transformation is used to form an abstraction to be applied to similar code.

LLMs have also been used for transformation by example recently. PyCraft \cite{dilhara2024unprecedented} uses a combination of static analysis, dynamic analysis, and LLMs with few shot learning to generate variants of input examples for conditional programmed attribute grammars. InferRules \cite{ketkar2022inferring} infers transformation rules from code examples in Java and Python.

CodePlan \cite{bairi2024codeplan} uses LLMs for code transformation, specifically focusing on repository-level changes using AI planning techniques and incremental dependency analysis to iteratively change code until it satisfies desired specifications. MELT \cite{ramos2023melt} utilizes GPT-4 to automatically generate transformation rules for Python libraries from code examples found in pull requests.

Another popular application of LLMs is for automating extract method refactoring \cite{fowler2018refactoring} which involves taking a section of code from a method and placing it in a new method. LM-Assist \cite{pomian2024together} combines LLMs with static analysis techniques and uses the IDE to perform refactoring. Revamp \cite{pailoor2024semantic} focuses on refactoring abstract data types using relational representation invariants to specify the desired changes to the data types.

Recent work explored using LLMs for automatic code migrations. Almeida et al. \cite{almeida2024automatic} investigates the effectiveness of using GPT 4.0 \cite{achiam2023gpt} to migrate eighteen methods and four tests in an application that utilized the SQLAlchemy library to a newer version. Tehrani and Anubhai \cite{omidvar2024evaluating} investigated human-AI partnerships for LLM-based code migrations by observing software developers working with Amazon Q Code Transformation to migrate systems from Java 7 to Java 18, and found that successful human-AI partnership relies on human-in-the-loop techniques.

These preliminary work are relevant to ours, though, to the best of our knowledge, our work is the largest, most comprehensive automated migration effort in scope and time frame to date.

\section{Conclusion and Future Work}

Code migration is a critical part of software development focused on updating and improving systems to ensure ongoing functionality. Manually handling code migrations, especially for large projects, is a challenging, time-consuming, and tedious process. Much research focused on automating migrations, where LLMs are a particular area of focus recently due to their abilities to manipulate code.

In this paper, we propose an automated solution for a large-scale, costly migration project at Google that was previously conducted manually. Our solution uses change location discovery and categorization, an LLM and automated validation to perform the migration automatically.

Our case study on \TotalIds{} distinct migrations over \TotalMonths{} months resulted in a total of \TotalCls{} code changes with \TotalEdits{} submitted edits. The LLM generated \TotalClsLlmPct{} of the submitted code changes, \TotalEditsLlmPct{} of the submitted edits, and developers using our system estimated the total time spent on the migration was reduced by \TotalTimeReduction{}.

Our results highlight the potential of LLMs to significantly improve developer productivity and efficiency in large-scale code migration tasks. By automating a substantial portion of the process, the system effectively minimized manual effort and saved developers considerable time. Developers highly appreciated the system's ability to accurately find potential code references, automate code changes, and validate modifications. They also noted that the continuous and automated nature of the system provided a clear sense of progress and reduced the tedious manual effort involved in code migration.

There were also challenges associated with the use of our system, namely the LLM's limited context window size and hallucinations. Additionally, the LLM's performance varied across different programming languages. Furthermore, pre-existing test failures hindered productivity and the largest migrations required safe production roll-outs.

Future research avenues include exploring ways to enhance LLM capabilities, such as increasing context window size, improving language support, mitigating hallucinations and enhancing the accuracy of code modifications, ultimately leading to improved software quality and developer productivity. Finally, a promising research direction is to use LLMs to not only assist in making code changes, but also to determine the code locations to be changed, potentially decreasing the number of left-over locations to be manually investigated.

\section{Acknowledgments}

This work is the result of a collaboration between the Google Core Developer, Google Ads and Google DeepMind teams. We thank key contributors: Priya Baliga, Siddharth Taneja, Ayoub Kachkach, Jonathan Bingham, Ballie Sandhu, Christoph Grotz, Alexander Frömmgen, Lera Kharatyan, Maxim Tabachnyk, Shardul Natu, Bar Karapetov, Kashmira Phalak, Andrew Villadsen, Maia Deutsch, AK Kulkarni, Satish Chandra, Danny Tarlow, Aditya Kini, Marc Brockschmidt, Yurun Shen, Milad Hashemi, Chris Gorgolewski, Don Schwarz, Chris Kennelly, Sarah Drasner, Niranjan Tulpule, Madhura Dudhgaonkar.

We also thank the internal Google reviewers, and the anonymous reviewers of the FSE committee.
\bibliographystyle{ACM-Reference-Format}
\balance
\bibliography{0-paper}

\end{document}